# Analysis of Influence of Internet Retail Service Quality (IRSQ) to Consumer Online Shopping Satisfaction at www.kebanaran.com

Imam Tahyudin
Department of Information System
STMIK AMIKOM PURWOKERTO
Purwokerto, Indonesia
imam.tahyudin@amikompurwokerto.ac.id

*Abstract* – The purpose of this research was to determine the influence of Internet Retail Service Quality (IRSQ) (website performance, access, security, sensation, and information) to the satisfaction www.kebanaran.com online shoppers. The method of analysis used was path analysis. Based on the research results influence IRSQ variables (performance, access, sensation, and information security), performance variables (X1), access (X2) and sensation (X3) had no significant effect on satisfaction (Y). It showsthat the online shopping website www.kebanaran.com already apply standard terms online stores in general, such as membership, has a return policy, a unique craft product offerings, the choice of language, the choice of currency, the chatroom facility, the product ctalogue about images from different angles and so forth, so that consumers be sure to purchase products through the online shopping website www.kebanaran.com. Security variable (X4) and information (X5) has a significant effect on satisfaction (Y). This shows that security is applied and the importance of information for consumers such as information availability, quality productsinformation, accurate product information is essential so that consumers do not hesitate to deal transaction use online shopping website www.kebanaran.com.

## I. INTRODUCTION

There are some experts who defines consumer satisfaction, satisfaction is the level of feelings after comparing the performance or results with the expectations [7]. According to Tse and Wilton, consumer satisfaction or dissatisfaction is the consumer response to the evaluation of the perceived discrepancy between prior expectations and actual performance of the product that is felt after consumption [15]. Wilkie define it as an emotional response to the evaluation of the experience of the product or service consumption [15]. Engel states that satisfaction is an evaluation after purchase as the chosen alternative at least equal or exceed consumer expectations, while dissatisfaction arise if the results do not meet expectations [15]. This suggests that to meet consumers' satisfaction are necessary to identify the consumers' expectations and then realize these expectations, so that consumers feel satisfied.

Among the companies are selling their products online. Higher internet penetration and the growing retail business that markets products and services by online demand the differences measurement of service quality between electronic retail services and the conventional services. Measuring the quality of services is intended to satisfy the consumer.

Quality of services centered on addressing the needs and wants of the consumer and delivery accuracy to offset consumer expectations [14]. According to Janda, Trocchia, and Gwinner [16] conducted research on consumer perceptions of Internet retail service quality and develop measurement scale that examined the quality of services from the perspective of the consumer. Measurement scale is organized into five main dimensions, namely: website performance, access, security of online shopping, shopping sensation, and information. The purpose of this research was to determine the influence of internet retail service quality (performance website, access, security, sensation, and information) to the consumer online shopping satisfaction at www.kebanaran.com.

## II. SERVICES QUALITY

The concept of services quality proposed by the Parasuraman et al. are relatively similar to the expectation paradigm which developed in the satisfaction research. In the research by Parasuraman et al measure consumer expectations for service company, these is consumers trust and the perception of reality regarding services received [14].

According Tjiptono stated that quality of services centered on addressing the needs and wants of the consumer and delivery accuracy to offset consumer expectations [14]. So there are two main factors that affect the quality of services, according to Parasuraman, namely expected service and perceived service. If the services received or perceived as expected, the perceived service quality and satisfactory [14]. If the services received exceed consumer expectations, the quality of service perceived as the ideal quality. Conversely, if the services received is lower than what is expected, then the perceived poor quality of services.

According to Zeithaml, consumer expectations for quality of a service is formed by:

a. *Enduring Service Intensifiers*
   This factor is a factor that is stable and encourage consumers to increase their sensitivity to services. This includes expectations caused by others and one's personal philosophy towards services.
b. *Personel Needs*
   One feels the need for fundamental welfare is also very decisive expectations. These needs include the physical, social and psychological.
c. *Transitory Service Intensifiers*
   This factor is a temporary individual factors that increase the sensitivity of consumers to the service, including:
   1) An emergency situation when a consumer really needs the services and wants the company can help.
   2) The consumer consumed the last service can also be a reference to determine the merits of subsequent services.
d. *Perceived Service Alternatives*
   It is the consumer's perception of the level or degree of service other similar companies. If consumers have few alternatives, the hopes for a service tends to be greater.
e. *Self Perceived Service Roles*
   This factor is the consumer's perception of the level or degree of involvement in influencing the services it receives.
f. Situational *Factors*
   This factor consists of all the possibilities that could influence the performance of services which are beyond the control of the service provider.
g. Explicit *Service Promises*
   This factor is the states of organization about their services to consumer. This promise as advertisment, personal selling, comunications with the employee.
h. Implicit *Service Promises*
   Regarding the instructions relating to services that allow consumers about the service and how it should be provided.
i. *Word of Mouth* (rekomendasi/saran dari orang lain)
   Is a statement made by someone other than the organization to consumers. Word of mouth is usually more readily accepted by the consumer, as are those that convey a credible as experts, friends, family and mass media publications.
j. *Past Experience*
   Past experience includes the things they have learned or known consumers from ever received in the past. This consumer expectations evolve over time, as more and more consumers as well as information received increasing numbers of consumer experience [14].

According Wyckof defined service quality is the level of excellence expected and control over the level of excellence to satisfy the consumer [13]. In defining the quality of service, there are some additional characteristics that should be considered. Garvin identified eight dimensions of quality, such as the performance characteristics of the operations on core products, features or additional privileges, compliance with specifications durability, serviceability, aesthetics and perception of quality [13]. However, most of the dimensions are more appropriately applied in manufacturing, therefore modified into seven dimensions that can be applied to service industries such as:

a. Function, the primary performance of the services required
b. Characteristics or additional features, the expected performance or complementary characteristics.
c. Conformance, a decision which is based on the fulfillment of specified conditions.
d. Reliability, belief in services in relation to time.
e. Serviceability, the ability to make repairs if there is some mistake.
f. Aesthetics, consumer experience associated with feelings and senses.
g. Perception, reputation for quality.

According to Janda, Trocchia, and Gwinner conducting research on consumer perceptions of Internet Retail Service Quality and develop measurement scale that examined the services quality from the perspective of the consumer. Measurement scale is organized into five main dimensions: website performance, access, security of online shopping, shopping sensation, and information [16].

### III. CUNSOMER SATISFACTION

In the services marketing literature, according to Bolton, Cronin and Taylor described consumer satisfaction as the decision on the basis of a specific service encounter [14]. This is accordance with Oliver'S [14] looked at satisfaction is an emotional reaction that affects attitude. From this perspective, Cronin and Taylor said that consumer satisfaction should confine the service quality and trading decisions specific to long-term attitudes. Consequently cumulative effect of service satisfaction needs to be directed at the evaluation of global service quality frequently [14]. Therefore, the researchers found that satisfaction was preceded by the quality of service [13]. Other researchers found the service quality and consumer satisfaction tested both a global perspective and the specific transaction [13]. According to Oliver and Goose satisfaction evaluation has been linked to the value [14], according to Kasper repeat purchase as well as consumer loyalty to the company [14]. Of course, Fornell in his study of Swedish consumers, which, although quality and consumer satisfaction is important for all companies, so satisfaction is more important than loyalty in the industry such as banking, insurance, postal orders, and transportation.

Many experts who provide a definition of consumer satisfaction. That consumer satisfaction or consumer dissatisfaction is the consumer response to the evaluation of discrepancy or disconfirmation perceived expectations previously (or other performance norms) and perceived actual performance of the product after its use [17].

Consumer satisfaction is an after-purchase evaluation alterantif selected where at least give the result (outcome) equal or exceed consumer expectations, while dissatisfaction arise if the results do not meet consumer expectations or in other words, consumer satisfaction is the behavior of one's feelings after comparing performance (or outcome) that he felt compared to his expectations [17].

Generally, cunsomer expectations are estimates or beliefs about what cunsomers would receive if bought or consuming a product (goods or services). While the perceived performance is the consumer's perception of what he received after consuming the products purchased [17].

According to Janda, Trocchia, and Gwinner conducting research on consumer perceptions of Internet Retail Service Quality and develop measurement scale that examined the quality of services from the perspective of the consumer [16]. Measurement scale is organized into five main dimensions: website performance, access, security of online shopping, shopping sensation, and information.

According to Garvin in evaluating product, service consumers satisfaction generally use multiple factors or dimensions, include [17]:
a. Performance
   That is the principal operating characteristics of the core product being purchased.
b. Features or additional privileges
   That secondary or complementary characteristics
c. Reliability
   That is unlikely to be damaged or fail to use
d. Conformance to specifications
   The extent to which the design and operating characteristics meet predetermined standards.
e. Durability
   With regard to how long the product can continue to be used.
f. Serviceability
   Includes speed, competence, comfort, easy to repair as well as a satisfactory complaint handling.
g. Aesthetics
   Product appeal to the five senses
h. Perceived quality
   That is the image and reputation of the product and the company's responsibility to it.

## IV. RESEARCH FINDINGS

### A. Path Analysis

To determine the influence of Internet retail service quality (performance website, access, security, sensation, and information) to the consumer online shopping satisfaction at www.kebanaran.com used path analysis. Having tested the validity and reliability of the data and then performed an ordinal transformation of data into interval data by the method of succesive interval (MSI). Once the data is converted into interval data and then do the path analysiscalculation. The results of path analysis calculations can be seen in table 1.

Table1. The results of path analysis calculations

| No | Variable | Path coef. | t - count | t -table | Sig. |
|----|----------|-----------|-----------|----------|------|
| 1 | Reality Performance | 0,138 | 1,615 | 1,9861 | 0,110 |
| 2 | Access reality | 0,003 | 0,030 | 1, 9861 | 0,976 |
| 3 | Reality sensation | -0,080 | -0,989 | 1, 9861 | 0,325 |
| 4 | Security reality | 0,275 | 3,005 | 1, 9861 | 0,003 |
| 5 | Reality information | 0,529 | 5,927 | 1, 9861 | 0,000 |
| Coefficient of determination = 0,564 | | | | | |
| F Count = 23,757 | | | | | |
| F table = 2,3134 | | | | | |

Total Influence of proportionally coefficient of determination ($R^2$) is 0.564, its mean that 56.40 percent of change in satisfaction can be explained by the variation of the variable performance, access, sensation, security and information services. Residual influence apart from the coefficient of determination is 0.4360, meaning that 43.60 percent is explained by other variables not examined.

### 1. F Test

To examine the path coefficients used F test to describes jointly influence the performance variable (X1), access (X2), sensation (X3), security (X4) and information (X5) satisfaction (Y). From calculating F test obtained F count is 23.757. Using the 95% significance

level($\alpha$=0,05) and degrees of freedom DF1 = (k-1) = 6-1 = 5, df2 = (n-k) = 98-6 = 92 obtained F table is 2.3134. So F count (23.757)> F table (2.3134) so Ho is rejected. Ho rejection means there is a significant influence of the variable performance, access, sensation, security and information services to satisfaction. The influence showed that the servicesquality provided by the website www.kebanaran.com been in line with expectations and satisfy the consumer.

The image of Ho rejection of F test can be seen in Figure 1.

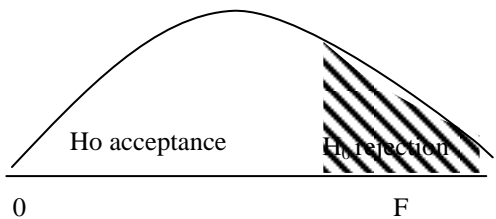

Figure1. Normal curveF test

**2. T test**

Using the 95% significance level($\alpha$ = 0,05) and degrees of freedom (n - k = 98-6 = 92) obtained t table is 1.9861, while the result of the calculation result t count performance variable (tYX1) is 1.615. So the t count is less than t table (1.615 <1.9861), which means that the partial performance variable (X1) has no significant impact on satisfaction (Y), so the first hypothesis is rejected. The lack of influence of performance on satisfaction showed that this online shopping website of used tires craft already apply standard provisions, such as the provision of a member willing to transact, have return policies, so that the consumer has no doubt to purchase through the online shopping webisiteat www.kebanaran.com.

tcount access variable (tYX2) is 0.03. So the t count is less than t table (0.003 <1.9861), which means that the partial access variable (X2) has no significant impact on consumer satisfaction (Y), so that the second hypothesis is rejected. The absence of these influenceshowed to date this online shopping website of used tires craft already apply standard provisions, such a unique craft product offerings, the choice of language, the choice of currency and so forth.

tcount sensationvariable (tYX3) is -0.989. So the t count is less than t table (-0.989 <-1.9861), its mean that the partial sensation variable (X3) had no significant impact on consumer satisfaction (Y), so that the third hypothesis is rejected. The absence of such influence because this online shopping website of used tires craft already apply standard provisions as a chatroom facility, product catalogue, so that consumers be convinced to buy used tires craft.

tcount security variable (tYX4) is 3.005. So the t count is greater than the value of t table (3.005> 1.9861), its mean that the partial security variables (X4) has a significant effect on cunsomer satisfaction (Y), so that the fourth hypothesis is accepted. The existence of thisinfuence suggests that security is applied in online shopping website does not make cunsomers hesitate to transact.

Value tcountof informationvariable (tYX5) is 5.927. So the t count is greater than the value of t table (5.927> 1.9861), mean that the partial information variable (X5) has a significant effect on consumer satisfaction (Y), so that the fifth hypothesis is accepted. The existence of thisinfluence suggest that the information presented on the website is very important to convince the consumer to buy the craft used tires product.

**V. CONCLUSIONS AND SUGGESTIONS**

**A. Conclusions**

Based on the research revealed that:
1. IRSQ variable (performance, access, sensation, information and security) jointly influentialto consumer online shopping satisfaction at www.kebanaran.com. Its obtained from the results of path analysis, F count (23.757) is greater than the F table (2.3134) so the hypothesis (H1) are received.
2. Performance variables (X1) had no significant influence on satisfaction (Y). This indicates that the craft tires online shopping website already used standard terms online shop in general, such as membership and return policy.
3. Access variable (X2) had no significant effect on satisfaction (Y). The absence of this influence that the craft tires online shopping website already used standard terms online shop in general as a unique craft product offerings, the choice of language and the choice of currency, so that cunsomers are no doubt to purchase

...

through the online shopping website www.kebanaran.com.

4. Sensation Variable (X3) had no significant influence on satisfaction (Y). The absence of thisinfluence shows craft tires online shopping website are standard terms used in general such as the online chatroom facility, the product can diliat images from different angles, so that consumers be sure to buy tires craft through the online shopping website www.kebanaran.com.

5. Security variable (X4) has a significant influence on satisfaction (Y). This shows that the security applied to the website www.kebanaran.com very important so that the consumers not hesitate to transact through the online shopping website.

6. Information variable (X5) has a significant influence on satisfaction (Y). The existence of this influence showed that the importance of information for consumers such as information availability, quality product information, product information is accurate.

B. Suggestions

1. Consumer satisfaction can be enhanced if the operator of the online shopping website further improve the services incorporated in the variables performance, access, sensation, security and information services. These variables should be increased simultaneously in order to optimize consumer satisfaction.

2. To optimize the consumer satisfaction, operator of the online shopping website (used tires craft) can better prioritize attributes incorporated in the performance variables and information. This is because the attributes are incorporated in the variable most dominant influence than other variables.

## ACKNOWLEDGMENT

The Authors would like to thank for the support and helpful comments of academicals member of STMIK AMIKOM Purwokerto for this work.## REFERENCES

[1]. Al Rasyid, 1994. Sampling Techniques and Preparation Scale. Program Pasca Sarjana Universitas Padjajaran Bandung.

[2]. Azwar, Saefudin. 2000. Test of Validity and Reliability. Pustaka Pelajar Yogyakarta

[3]. Caruana, Albert.2002. The effects of service quality and the mediating role of consumer satisfaction.*European Journal of Marketing Volume 36 Number 7/8 2002 pp. 811-828.*

[4]. Cooper dan Emory, 1998. Business Research Methods. Erlangga Publisher Jakarta.

[5]. Hazlina Abdul Kadir, Nasim Rahmani and Reza Masinaei. 2011. Impacts of service quality on consumer satisfaction: Study of Online banking and ATM services in Malaysia. International Journal of Trade, Economics and Finance, Vol.2, No.1.

[6]. Jayaraman Munusamy, Shankar Chelliah and Hor Wai Mun. 2010. Service Quality Delivery and Its Impact on Consumer Satisfaction in the Banking Sector in Malaysia. International Journal of Innovation, Management and Technology, Vol. 1, No. 4.

[7]. Kotler, Philip. 2001. Marketing Management (Book 2). Jakarta: Salemba Empat.

[8]. Mowen, J.C. 1995. Consumer Behavior. 4th edition. Prentice Hall Inc New Jersey. Nha Nguyen dan Gaston LeBlanc. 1998. The mediating role of corporate image on consumers' retention decisions: an investigation in financial services.International Journal of Bank Marketing Volume 16 Number 2 1998 pp. 52-65

[9]. Prasetyo Adi. 2008. Influence Analysis Of Service Quality Customer Satisfaction Kaffah BMT Yogyakarta. STAIN Surakarta SEM Institute, Yogyakarta.

[10]. Rambat Lupiyoadi. 2004. Marketing Management Services : Teory and implementation. Jakarta: PT salemba Empat.

[11]. ---------------------- dan A. Hamdani. 2006. Marketing Management Services. Jakarta: Salemba Empat.

[12]. Sitepu, Nirwana SK. 1994. Path Analysis. Program PascaSarjana Universitas Padjajaran Bandung.

[13]. Tjiptono, Fandy. 1998. Management Services. Andy Publisher Yogyakarta

[14]. -------------------. 2000. The principles of Total Quality Service, First Printing, Second Edition, Andi Publisher, Yogyakarta.

[15]. -------------------, 2001. Marketing Strategy. Andi Publisher. Yogyakarta.

[16]. -------------------, Chandra Yanto, Diana Anastasia. 2003. Marketing Scale. Andi Publisher. Yogyakarta.

[17]. -------------------. 2007. Service Marketing. Bayumedia Publishing, Malang.

[18]. Umar, Husein 2000. Conduct Market Research and Marketing. Gramedia Publisher, Pustaka Utama Jakarta.

## AUTHORS PROFILE

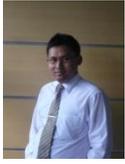Imam Tahyudin was born in Indramayu, West Java, Indonesia, on July 12, 1983. He Received B.Sc. degree from Faculty of Science and Technology, Universitas Jenderal Soedirman Purwokerto, Indonesia in 2006 and M.M. degree from faculty of Economic Universitas Jenderal Soedirman Purwokerto, Indonesia in 2010. He is currently pursuing the M.Eng. degree in the department of Information Engineering, STMIK AMIKOM Yogyakarta, Indonesia, in the field of information system. He is lecturer in the department of information system STMIK AMIKOM Purwokerto, Indonesia. His research interests are in information system management and data mining.